\documentclass[structabstract]{aa}

\usepackage{graphicx}
\usepackage{txfonts}
\usepackage{natbib}
\usepackage{longtable}
\bibpunct{(}{)}{;}{a}{}{,}

\begin{document}
\title{Infrared-Faint Radio Sources: A Cosmological View}

\subtitle{AGN Number Counts, the Cosmic X-Ray
  Background and SMBH Formation}

\author{Peter-Christian Zinn
  \inst{1}
  \and
  Enno Middelberg
  \inst{1}
  \and
  Edo Ibar
  \inst{2}
}

\institute{Astronomical Institute, Ruhr-University Bochum, Universit\"atsstra\ss{}e 150, D-44801 Bochum\\
  \email{zinn@astro.rub.de}
  \and UK Astronomy Technology Centre, Royal Observatory, Blackford Hill, Edinburgh EH9 3HJ, UK
}
\date{Received 06/12/2010; accepted 31/03/2011}

\abstract {Infrared Faint Radio Sources (IFRS)
  are extragalactic emitters clearly detected at
  radio wavelengths but barely detected or undetected at optical
  and infrared wavelengths, with 5$\sigma$ sensitivities as low as 1\,$\mu$Jy.} {Recent SED-modelling and
  analysis of their radio properties shows that
  IFRS are consistent with a population of (potentially extremely obscured) high-redshift AGN
  at 3\,$\leq$\,z\,$\leq$\,6. We
  demonstrate some astrophysical implications of this population and
  compare them to predictions from models of galaxy evolution and
  structure formation.} {We compiled a list of IFRS from four deep
  extragalactic surveys and extrapolated the IFRS number density to a
  survey-independent value of (30.8$\pm$15.0)\,deg$^{-2}$. We computed
  the IFRS contribution to the total number of AGN in the Universe
  to account for the Cosmic X-ray Background. By estimating the
  black hole mass contained in IFRS, we present conclusions for
  the SMBH mass density in the early universe and compare it to
  relevant simulations of structure formation after the Big Bang.}{The
  number density of AGN derived from the IFRS density
  was found to be $\sim$\,310\,deg$^{-2}$, which is equivalent to a
  SMBH mass density of the order of $10^3$\,M$_{\sun}$\,Mpc$^{-3}$ in the redshift range 3\,$\leq$\,z\,$\leq$\,6. This produces an X-ray flux of $9\cdot
  10^{-16}$\,W\,m$^{-2}$\,deg$^{-2}$ in the 0.5-2.0\,keV
  band and $3\cdot 10^{-15}$\,W\,m$^{-2}$\,deg$^{-2}$ in
  the 2.0-10\,keV band, in agreement with the
  missing unresolved components of the Cosmic X-ray
  Background. Concerning the problem of SMBH formation after the Big
  Bang we find evidence for a scenario involving both halo gas
  accretion and major mergers.}  {}


\maketitle

\section{Introduction}
\label{intro}
Infrared-Faint Radio Sources (IFRS) were first classified by
\cite{Norris2006}, who identified them as sources detected at 1.4\,GHz
using the the Australia Telescope Compact Array (ATCA), but absent in
deep infrared images from the {\it Spitzer} Wide-Area Infrared
  Extragalactic (SWIRE) survey (\citealt{Lonsdale2003}). In the
  process of matching IR counterparts to radio sources from the
  Australia Telescope Large Area Survey (ATLAS),
  \citeauthor{Norris2006} found a total of 22 radio sources to which
  no infrared counterpart could be reasonably identified in any of the
  observed {\it Spitzer} bands (3.6\,$\mu$m -- 24\,$\mu$m), down to a
  sensitivity of $\sigma_{3.6\,\mu m}\,=\,1.0\,\mu$Jy. A similar
result was obtained by \cite{Middelberg2008a} who found 31 IFRS in the
ELAIS-S1 at similar IR and radio sensitivities. A substantial fraction
of these IFRS have flux densities of several mJy at 1.4\,GHz, and some
are even found at 20\,mJy.

A first attempt to investigate the nature of the IFRS discovered in
the ATLAS survey was undertaken by \cite{Norris2007}, who
observed two IFRS with Very Long Baseline Interferometry (VLBI), and
detected one. They argued that IFRS are AGN-driven objects
  because the VLBI detection indicated a brightness temperatures in
  excess of $10^6$\,K. Such high temperatures cannot be reached by
  thermal emission mechanisms and hence indicate non-thermal radio
  emission produced by an AGN. Later \cite{Middelberg2008b}, also
using VLBI observations, confirmed that at least a fraction of the
IFRS population must contain an AGN contributing to the total IFRS
spectral energy distribution.

These results were complemented by \cite{Garn2008} who
identified an IFRS population in the {\it Spitzer} extragalactic First
Look Survey (xFLS). Modelling their spectral energy distributions
(SEDs) with template SEDs from known objects to estimate IFRS
redshifts, they found that the characteristics of IFRS resembled a
variety of 3C sources redshifted to 2\,$\leq$\,z\,$\leq$\,5. A more
detailed SED analysis was presented by \cite{Huynh2010}, who
used new, very deep {\it Spitzer} data in the {\it Chandra}
  Deep Field South (CDF-S) field obtained during two {\it Spitzer}
legacy programs, SIMPLE (Damen et al., in prep.)  and FIDEL (PI:
Dickinson). With a sensitivity of $\sigma_{3.6\,\rm{\mu
    m}}\,=\,0.16\,\mu$Jy, IR counterparts for two of the four
  investigated IFRS were identified. Furthermore, a faint
  optical counterpart (V\,=\,26.27\,AB-mag) to one of the IFRS was
  serendipitously identified within the GOODS HST/ACS imaging campaign
  \citep{Giavalisco2004}.  A second IFRS located within the
  GOODS-South field was found to have no optical detection at all
  (implying V\,$>$\,28.9\,AB-mag), putting strong constraints on their
  SEDs. Huynh et al. also searched other deep multi-wavelengths data
  sets in the CDF-S for counterparts, in particular the {\it Chandra}
  2\,Ms source catalog \citep{Luo2008}, the GEMS HST imaging campaign
  \citep{Rix2004} and the MUSYC catalog \citep{Gawiser2006}. These
  searches did not yield any counterparts in the X-ray or the
  optical/NIR regime. Modelling template SEDs to represent the radio
and, if available, the IR and optical detections, \cite{Huynh2010}
found that a 3C\,273-like object can reproduce the data when
redshifted to z\,=\,2 (IR-detection) and z\,$>$\,4 (IR
non-detection). The IR non-detections could not be explained by any
template SED at redshifts smaller than $\sim$4. These estimates
constrain the redshift range of IFRS, placing at least a significant
fraction of them at z\,$>$\,4. On the other hand, given that none of
the IFRS was detected at 24\,$\mu$m, \cite{Huynh2010} conclude that
IFRS fall well beyond the IR-radio correlation \citep[see e.g.][whose
  $q_{24}\,=\,\log\left(S_{24\,\rm{\mu m}}/S_{20\,\rm{cm}}\right)$
  parameter of 0.84 would imply a 24\,$\mu$m flux density of 7\,mJy
  for a typical IFRS with
  S$_{20\,\rm{cm}}\,=\,1\,$mJy]{Appleton2004}. Hence the IFRS
radio emission can not be explained by star-forming processes but
  is likely to be produced by AGN. This finding is supported by
applying various calibrations of the 1.4\,GHz luminosity as star
formation rate (SFR) tracer. For instance, the classical calibration
by \cite{Bell2003} gives star formation rates of around a million
solar masses per year for a typical IFRS assumed to be located at
z\,=\,4, which is unphysical and hence suggests that the radio
  emission of IFRS cannot be only caused by star formation.

Using new radio data spanning 2.3\,GHz to 8.6\,GHz,
\cite{Middelberg2011} analysed the radio properties of 18 ATLAS IFRS.
They found unusually steep radio spectral indices with a median
$\alpha$ of $-1.4$ ($S\propto\nu^\alpha$), and no spectral index
flatter than $-0.5$. According to the z--$\alpha$ relation
\citep[e.g.][]{Athreya1998} that predicts steeper radio spectral
indices at higher redshifts, these findings suggest that IFRS reside
in the high-redshift Universe. Another result from Middelberg et
al. is that the ratio between radio and IR flux densities,
$S_{20\,\rm{cm}}/S_{3.6\,\rm{\mu m}}$, is much larger for IFRS than
for the general radio source population, and potentially exceeds
$10^4$, a value which is similar to that found in a sample of
high redshift radio galaxies (HzRGs) by \cite{Seymour2007}. These
ratios and redshift dependencies are further investigated in more
detail by \cite{Norris2011}. They suggest that IFRS have redshifts of
z\,$\sim$\,5 in the most extreme cases, because no 3.6\,$\mu$m
counterpart could be identified by stacking ultra-deep {\it
  Spitzer} observations at the location of 39 IFRS, reaching a
  stacked sensitivity of $S_{\rm 3.6\,\mu m}\sim$0.2\,$\mu$Jy.

In summary, the evidence from SED modelling, very high
$S_{20\,\rm{cm}}/S_{3.6\,\rm{\mu m}}$ ratios and steep spectral
indices all suggest that IFRS are at high redshifts (at least z\,$>$\,2) and contain
AGN. In this paper we present the implications of this population for
the AGN number densities, the Cosmic X-ray background (CXB), and
structure formation models. Throughout this paper, we adopt a standard
flat $\Lambda$CDM cosmology with
$H_0\,=\,71$\,km\,s$^{-1}$\,Mpc$^{-1}$ and $\Omega_M$\,=\,0.27.

\section{Catalogue of known IFRS}

In previous publications the selection criterion used to identify IFRS
was simply ``radio sources with no apparent IR counterpart''. This is
a rather loose definition, and we propose to replace this definition
by the following two criteria:

\begin{itemize}
\item The ratio of radio to mid-IR flux,
  $S_{20\,\rm{cm}}/S_{3.6\,\rm{\mu m}}$, exceeds 500.
\item The 3.6\,$\mu$m flux density is smaller than 30\,$\mu$Jy.
\end{itemize}

The first criterion selects objects with extreme radio to IR flux
ratios, making IFRS clear outliers from the IR-radio correlation. This
encapsulates the requirement that IFRS are strong in the radio, but
weak at infrared wavelengths. The second criterion ensures that the
sources are at significant, cosmologically relevant, distances, to
eliminate low-redshift radio-loud AGNs. For example, Cygnus~A (with
$S_{20\,\rm{cm}}/S_{3.6\,\rm{\mu m}}$\,$=$\,2$\cdot$10$^5$) would be
an IFRS based on its radio to IR flux ratio alone, but would be
excluded by the 3.6\,$\mu$m flux density cutoff.

\cite{Alonso-Herrero2006} have shown that the near-to-mid-IR SEDs
  of galaxies dominated by AGN emission exhibit a power-law behaviour
produced by re-emission from warm dust surrounding the AGN. They find
that AGN can span a large range of optical-to-IR spectral slopes
with a maximum $\alpha=-2.8$ but typically between $-0.5$ and
$-1.0$. These values are similar to those observed in the
radio. In fact, the k-corrections necessary to estimate the intrinsic
$S_{20\,\rm{cm}}/S_{3.6\,\rm{\mu m}}$ values would produce a small
difference with respect to the observed ones. It is evident that the
first criterion preferably select IFRS if the spectral slope in
the near-IR is steeper than in radio, i.e.\ star-forming galaxies
(with flatter spectra due to a dominant stellar emission) are
  preferrably scattered away from the sample. This weak dependency
for $S_{20\,\rm{cm}}/S_{3.6\,\rm{\mu m}}$ as a function of redshift
suggests that IFRS are intrinsically extreme outliers of the IR-radio
correlation. This leaves the second criterion as the dominant to
preferentially select high-redshift sources using the dimming produced
by cosmic distance.

Applying our criteria, we present in this paper a catalogue containing
all IFRS found in the ATLAS/CDF-S \citep{Norris2006} and
ATLAS/ELAIS-S1 \citep{Middelberg2008a} surveys, in the {\it Spitzer}
extragalactic First Look Survey \citep[xFLS,][]{Condon2003}, and in
the COSMOS survey \citep{Schinnerer2007}. To date, this compilation is
the most comprehensive list of IFRS, including 55 objects located in
four individual surveys.  The surveys were chosen not only
  because of their radio and IR coverage but also for the covered
area. To minimise the effects of cosmic variance, only surveys greater
than 1\,deg$^2$ were analysed. Cosmic variance can be a significant source of error when measuring density-related
quantities especially in small survey fields. For example, the
GOODS field was analysed for cosmic variance by
\cite{Somerville2004}. They found that for fields in the order of
several 100\,arcmin$^2$, such as GOODS, the error of number density
estimates can range from 20\% up to 60\% for strongly clustered
sources. To avoid statistical errors introduced by cosmic variance in
the estimated number counts, we have excluded small area surveys. In
particular we have left out the deep surveys in the Lockman Hole East
(\citealt{Ibar2009}) and Lockman Hole North (\citealt{Owen2008}). An
overview of the surveys we used is shown in Table \ref{surveys}.

\begin{table}[t]
  \caption{Summary of investigated surveys.}
  \label{surveys}
  \centering
  \begin{tabular}{lcccc}
    \hline
    \hline
    \noalign{\smallskip}
    Survey&Area&3.6\,$\mu$m limit&20\,cm limit&IFRS\\
    Name&deg$^2$&$\mu$Jy&$\mu$Jy&number\\
    \noalign{\smallskip}
    \hline
    \noalign{\smallskip}
    ATLAS/CDF-S\tablefootmark{a}&3.7&3.1&186&14\\
    ATLAS/ELAIS-S1\tablefootmark{a}&3.6&3.1&160&15\\
    First Look Survey\tablefootmark{b}&3.1&9.0&105&13\\
    COSMOS\tablefootmark{c}&1.1&1.0&65&13\\
    \noalign{\smallskip}
    \hline
  \end{tabular}
  \tablefoot{The given sensitivity limits are all $5\sigma$.\\
    \tablefoottext{a}{Radio from \cite{Norris2006} and \cite{Middelberg2008a}, IR from \cite{Lonsdale2003}.}\\
    \tablefoottext{b}{Radio from \cite{Condon2003} and IR from \cite{Lacy2005}.}\\
    \tablefoottext{c}{Radio from \cite{Schinnerer2007} (only the inner part of the 20\,cm map was searched for IFRS to assure a homogeneous noise level across the searched area) and IR from \cite{Sanders2007}.}
  }
\end{table}

\begin{figure}[t]
  \centering
  \includegraphics[width=0.48\textwidth,clip]{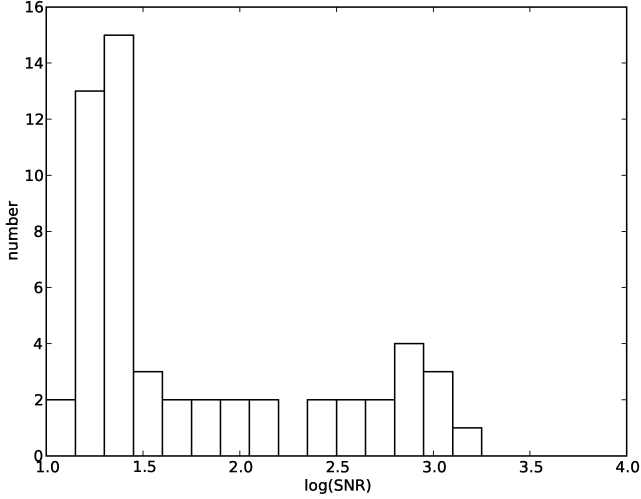}
  \caption{Histogram of the SNR distribution at 20\,cm for all IFRS
    presented in this paper. Note that the abscissa is logarithmic.}
  \label{IFRS-SNR}
\end{figure}

The source selection process was as follows: a pre-selection of
candidate sources was obtained by cross-matching radio and IR
catalogues of each survey in two different ways. Clearly extended
radio sources, or those consisting of multiple components, were
excluded from this pre-selection to prevent resolved AGN lobes
entering the sample. All radio sources with a 3.6\,$\mu$m counterpart
within a 5$\arcsec$  radius were selected and their
$S_{20\,\rm{cm}}/S_{3.6\,\rm{\mu m}}$ ratio was calculated. For all
radio sources that did not have a 3.6\,$\mu$m counterpart within
5\arcsec, the 5$\sigma$ detection limit of each survey as listed in
Table~\ref{surveys} was chosen to provide a conservative upper limit
on the 3.6\,$\mu$m flux density. Hence the inferred
$S_{20\,\rm{cm}}/S_{3.6\,\rm{\mu m}}$ is a lower limit, and is
indicated as such in Table~\ref{IFRS-cat}. A 5$\arcsec$  search radius
for cross-matching the radio and 3.6\,$\mu$m data ensures that at any
redshift greater than z\,=\,0.5, the corresponding linear
distance is on the order of 30\,kpc or more, and so exceeds the
size of a large spiral galaxy. We then produced postage stamp images
for visual inspection of each candidate source, showing a grey-scale
IR image centred on the radio position of the source with overlaid
radio contours. These images were inspected by eye to determine
whether the source really is an IFRS or whether the cross-matching
process failed, e.g. because of an uncatalogued IR counterpart.
  We also addressed the problem of source blending due to the
  different resolutions of the IR and radio images (FWHM of
  1.66$\arcsec$  in the IRAC $3.6\,\rm{\mu m}$ channel, \citealt{Fazio2004},
  vs. 11$\arcsec$  by 5$\arcsec$  in the ATLAS 1.4\,GHz radio data). IFRS
  candidates with more than one plausible IR counterpart were rejected
  in the selection process, since $S_{20\,\rm{cm}}/S_{3.6\,\rm{\mu
      m}}$ could not be reliably determined. The final sample
of 55 sources is listed in Table~\ref{IFRS-cat}, and a histogram
showing the distribution of radio SNR is shown in
Figure~\ref{IFRS-SNR}.

Note that the visual inspection of candidate sources helped to
  keep the rate of false detections (i.e., arising from imaging
artefacts) to a minimum. The survey sensitivities in
Table~\ref{surveys} were obtained by calculating the median of the rms
at the source positions, as provided in the respective catalogues. We
used this as characteristic survey sensitivity rather than adopting
the sensitivity limits given in the relevant publications because
those limits were often related to the most sensitive regions of the
observed areas, and thus were not representative.

\section{IFRS Densities}

\subsection{The sample}

To derive an estimate of IFRS surface density, at least to within
  an order of magnitude, we analysed the four surveys quoted
above. We used surveys with different sensitivities to
  extrapolate to the total surface density of IFRS in the Universe.

As shown in Table \ref{surveys}, the surface density of IFRS strongly
depends on the sensitivity of the individual radio surveys, whereas
the IR sensitivity seems to have little effect. The errors for the
individual measurements were obtained as follows: the error in radio
sensitivity is the Median Absolute Deviation (MAD) of the rms values
at the position of the targets from a given radio survey. This measure
is robust against outliers and can characterise non-Gaussian
distributions. Assigning errors to the measured IFRS surface densities
is more complicated. We chose to follow a statistical {\it ansatz}
claiming that there is an intrinsic IFRS surface density at the
sensitivity limit of a given survey, $\lambda_{intr}$, and the IFRS
surface density that we actually measured for that survey,
$\lambda_{meas}$. Assuming a Poissonian distribution (because the IFRS
surface density is always greater than zero and discrete), we
determined the 95\,\% confidence levels with which one would measure a
surface density of $\lambda_{meas}$, given the true surface density is
$\lambda_{intr}$. Because the Poisson distribution is asymmetric, this
procedure yielded asymmetric errors for the surface densities.

\begin{figure}[t]
  \centering
  \includegraphics[width=0.48\textwidth,clip]{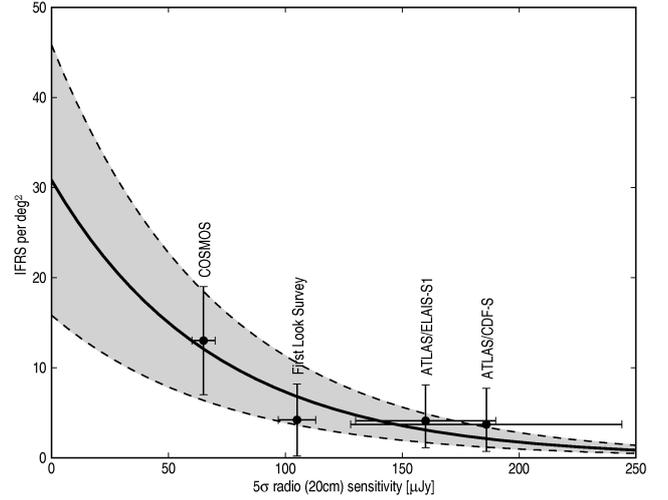}
  \caption{Plot illustrating the extrapolation of IFRS surface density
    with respect to different radio sensitivities. The best fit (solid
    line) is given by $\lambda\,=\,(30.8\pm 15.0)\exp\{(-0.014\pm 0.006)5\sigma_{radio}\}$,
    its quality in terms of $\chi^2/d.o.f.$ is 5.5, Pearson's correlation coefficient is $R^2\,=\,0.82$. The dashed curves that limit the shaded area correspond to fit functions representing the minimum and maximum values for the IFRS surface density with respect to the error or $\pm$\,15.0 given above.}
  \label{IFRS-dens}
\end{figure}

To extrapolate the measured densities to an intrinsic IFRS surface
density (what a radio observation with zero noise would yield), an
exponential fit was used to model the measurements (shown in
Fig.~\ref{IFRS-dens}). We chose an exponential function because
of a theoretical and a practical reason. The theoretical reason is
that the function should be monotonic because there is no obvious
reason why the IFRS surface density should peak around a certain
sensitivity. The practical reason is that in our tests an
exponential function yielded the best results with respect
to a $\chi^2$ test ($\chi^2/d.o.f.$\,=\,5.5). We thus define the
extrapolated IFRS surface density as the value given by the y-axis
intersect of the fit at $(30.8\pm 15.0)$\,IFRS/deg$^2$. We point
  out that this large error is real, underlining the fact that this
  calculation has to be understood as ``order of magnitude'' estimate
  only.

\subsection{Motivation of a redshift range for IFRS}

To convert the surface density of IFRS to a space density we assume
that the bulk of IFRS populate the redshift range
3\,$\leq$\,z\,$\leq$\,6. There are several motivations for this
choice: (i) it has been suggested by the SED modelling work described
in Section~\ref{intro} that most IFRS SEDs are inconsistent with
template SEDs displaced to redshifts significantly less than z\,=\,4;
(ii) those few IFRS detected in 3.6$\mu$m with this ultra-deep IR data
are at least at z\,=\,2 (conclusions in \citealt{Huynh2010}); (iii)
the radio to IR flux density ratios of IFRS are similar to that of the
\cite{Seymour2007} HzRG sample, of which many sources have redshifts
of 3 and above; (iv) IFRS have very steep radio spectra, which
indicate high redshifts according to the z--$\alpha$ relation.
  However, since there is a large intrinsic scatter in the
  z--$\alpha$ relation as e.g. given by \cite{Athreya1998}, we do not
  give quantitative redshift estimates based on this relation but only
  employ it as a qualitative hint. Therefore we adopt z\,=\,3 to be a
good compromise as lower limit for their redshift range.

The upper limit of z\,=\,6 was chosen because of the scarcity of
sources beyond this redshift, as has been found, e.g., in the Sloan
Digital Sky Survey \citep[SDSS; see
e.g.][]{Schneider2010,Fan2003}. Whilst there is a significant number
of quasars at redshifts of z\,$\sim$\,5, their number rapidly
decreases beyond z\,$\sim$\,6, with only very few spectroscopically
confirmed quasars having redshifts significantly greater than 6. Hence
z\,=\,6 is a reasonable upper limit for our estimates. The problem of
obtaining redshift estimates for IFRS is discussed in more detail by
\cite{Norris2011}, who argue for redshifts as high as z\,=\,5,
using extrapolations of the $S_{20\,\rm{cm}}/S_{3.6\,\rm{\mu m}}$
ratios and 3.6\,$\mu$m flux densities of typical IFRS.

\begin{figure}[t]
  \centering
  \includegraphics[width=0.48\textwidth,clip]{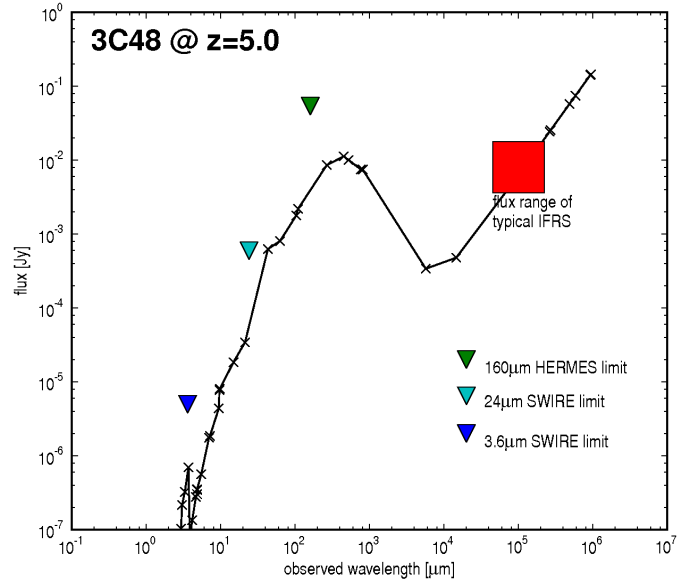}
  \caption{The template SED of 3C\,48, redshifted to z\,=\,5, boosted
    in flux by a wavelength-independent factor of 5 and dimmed in the
    optical and NIR following a \cite{Calzetti2000} extinction law
    with $A_V\,=\,1$. To emphasise that this SED is the one of an
    IFRS, we give the detection limits of the SWIRE survey described
    in sect. \ref{intro} as well as the 160\,$\mu$m limit of the
    planned {\it Herschel} Hermes survey. The radio spectral index of
    this SED corresponds to $\alpha\,=\,-1.4$.}
  \label{3C48}
\end{figure}

Furthermore, to also get a theoretically motivated estimate of the
minimum redshift for a typical IFRS, we extended the work by
\cite{Huynh2010} who modelled the SEDs of IFRS by known template
SEDs of different astrophysical objects (see Sec.~\ref{intro}). We
chose 3C\,48 with an intrinsic redshift of 0.37 as
template. 3C\,48 is an excellent candidate for being an IFRS at higher
redshifts because of its intrinsic high
$S_{20\,\rm{cm}}/S_{3.6\,\rm{\mu m}}$ ratio; it therefore does
not require huge amounts of obscuring dust to dim down the
observed NIR. We redshifted it to z\,=\,5, boosted it in flux by
a constant factor of 5 and dimmed it in the optical in NIR wavelength
range using a \cite{Calzetti2000} reddening law with $A_V\,=\,1$. The
resulting model SED can reproduce all observational characteristics of
a typical IFRS. This result is illustrated in Fig. \ref{3C48}.

Using this model SED, we traced the evolution of its
$S_{20\,\rm{cm}}/S_{3.6\,\rm{\mu m}}$ flux ratio as a function of
redshift. We also included several radio spectral indices in our
calculations to account for different radio components (e.g. bremsstrahlung, aged synchrotron or flat AGN spectra) which could yield different radio spectral indices than the canonical
$\alpha\,=\,-0.7$ ($S_{\nu}\propto\nu^{\alpha}$) for thermal
synchrotron emission. Thus we account for the radio spectral
indices being typically lower towards higher redshift.

\begin{figure}[t]
  \centering
  \includegraphics[width=0.48\textwidth,clip]{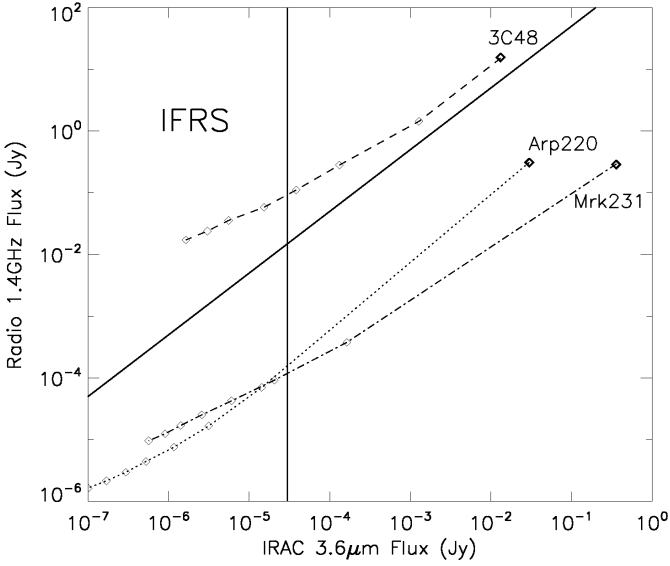}
  \caption{Tracks showing the evolution of the radio and NIR
    luminosity of a 3C\,48-like model with redshift. The redshift data
    points are given from right to left, starting with 3C\,48's (and
    the other two comparative sources, Mrk\,231 and Arp\,220)
    intrinsic redshift (bold open square). The first thin open square
    from right indicates a redshift of z\,=\,1, the next thin open
    square is equal to z\,=\,2 and so on to z\,=\,7. The vertical
    solid line indicates the IFRS NIR flux cutoff-criterion
    $S_{3.6\,\rm{\mu m}}\,<\,30\,\mu$Jy, the other solid line
    represents $S_{20\,\rm{cm}}/S_{3.6\,\rm{\mu m}}\,=\,500$.}
  \label{model}
\end{figure}

The resulting plot (Fig. \ref{model}) shows that only beyond a
redshift of $z=3$ 3C\,48-like models are able to reproduce the high
$S_{20\,\rm{cm}}/S_{3.6\,\rm{\mu m}}$ flux ratios and the low
3.6\,$\mu$m flux densities observed in IFRS. Our model calculations
indicate that to reproduce the observed 1.4\,GHz flux densities of a
few mJy, steep spectral indices are preferred, which is in agreement
with the results obtained by \cite{Middelberg2011}.

\subsection{The space density of IFRS and their SMBH}

In our redshift range of $z=3-6$, a solid angle of 1\,deg$^2$
corresponds to a comoving volume of $21.5\cdot 10^{6}$\,Mpc$^3$
\citep{Wright2006}, and so the observed surface density of IFRS
  converts to an IFRS space density simply by dividing the
extrapolated number density of IFRS per deg$^2$ by the comoving
volume:

\begin{equation}
  n_{IFRS}\,=\,1.44\cdot 10^{-6}\,\rm{Mpc}^{-3}
\end{equation}

This space density can now be converted into a mass density of
Supermassive Black Holes (SMBHs) that power IFRS. However, simply
adopting SMBH masses found in recent studies of large quasar surveys
as described in \cite{Vestergaard2008} for the SDSS data release 3
(containing more than 15\,000 quasars up to z\,=\,5) would lead to an
overestimation of SMBH masses because the limits of contemporary
optical surveys prevent from detecting high-z quasars with less than
$M_{SMBH}\,\sim\,10^8\,$M$_{\sun}$. The actual limit for the SDSS is
about $3\cdot 10^8\,$M$_{\sun}$ and is based on the line width of
broad emission lines in a quasar spectra to determine $M_{SMBH}$ from
their equivalent width. When converted to optical luminosities, this
corresponds to a lower limit of $M_i\,=\,-27.5$ for a z\,=\,4
quasar. IFRS cannot be optically bright quasars, because this would
require large amounts of dust in their surrounding in order to
maintain their IFRS nature, hence one has to lower the average
SMBH mass found by \cite{Vestergaard2008,Vestergaard2009} for
high-redshift AGN of $\langle M_{SMBH}\rangle\,\sim\,10^9\,$M$_{\sun}$
\citep[this value is supported by other work with much smaller quasar
  samples but more accurate methods for determining $M_{SMBH}$, see
  e.g. ][]{Dietrich2004,Fine2008}. For example, an SMBH radiating at
its Eddington limit produces an unobscured z\,=\,4 quasar with an
apparent magnitude of 25.0 (a common limit for contemporary optical
surveys carried out with space-based facilities) while having a mass
of only $3.5\cdot 10^6\,$M$_{\sun}$. Recent cosmological
hydrodynamical simulations \citep[e.g. ][]{Degraf2010} showed that
most SMBHs at z\,=\,5 do not accrete at their Eddington limit but at a
factor of approximately 5 to 10 times less. Nevertheless,
observational evidence for the $L_{Bol}/L_{Edd}$ ratio indicates an
accretion mechanism at $L_{Bol}/L_{Edd}\,=\,1$ for the most distant
quasars \citep{Fan2009}. This discrepancy between theory and
observations could be a selection effect in currently available
observations because SMBHs not accreting at their Eddington limit
would simply be too faint to be detected. Therefore we conclude that
optical observations are unsuitable to estimate a typical SMBH mass
for IFRS.

Another way to estimate the SMBH masses in IFRS are X-ray observations
which are thought to be an efficient way to detect large numbers of
AGN. X-rays can leave the emission regions rather unabsorbed, and
there is little contamination from other sources, such as stars. Given
that no IFRS identified in the radio observations of the ELAIS-S1
field is detected in the corresponding X-ray survey by
\cite{Puccetti2006}, we can use this information to derive an upper
limit for a SMBH. Using Equation \ref{eq:SSMBH}, a $4\cdot 10^7\,{\rm
  M}_{\sun}$ SMBH at z\,=\,4.5 would just not be visible in the
0.5-2.0\,keV band, given the detection limits of the
\cite{Puccetti2006} survey of
5.5$\cdot$10$^{-19}$\,W\,m$^{-2}$\footnote{For the convenience of the
  reader we give the conversion factor between SI and old cgs units:
  1\,W\,m$^{-2}\,\equiv\,10^3\,\rm{erg}\,\rm{s}^{-1}\,\rm{cm}^{-2}$}. Unfortunately,
no IFRS is located in the CDF-S {\it Chandra} 2\,Ms observation by
\cite{Luo2008}. The X-ray observations done within the COSMOS survey
using XMM-{\it Newton} \citep[best sensitivity of
  7.2$\cdot$10$^{-19}$\,W\,m$^{-2}$ in the soft band]{Cappelluti2007}
did not yield a clear identification of an IFRS X-ray counterpart,
too. We note that especially for the inner parts of the COSMOS X-ray
image, the field is very crowded which makes the reliable
detection of potentially very weak counterparts
  difficult. However, to provide some margin (to account for
extinction, for example), we approximately double this value and
therefore assume an {\it upper limit} for the mass for SMBHs in IFRS
of $M_{IFRS}\,\leq\,10^8\,$M$_{\sun}$. Hence the mass density of SMBH
in IFRS is:

\begin{equation}
  \rho_{IFRS}\,\leq\,1.4\cdot 10^2\,\rm{M_{\sun}\,Mpc}^{-3}
\end{equation}

\section{Implications}
\subsection{... on AGN Source Counts}

The hypothesis that IFRS contain AGN opens the opportunity to further
constrain the total AGN number density and to estimate their
contribution to the total AGN population. To do that, one has to
correct for not all AGN being radio sources, i.e., there is a
population of faint AGN (due to an intrinsically low luminosity or
because they are obscured by dust) of which we can identify only a
fraction because of their radio emission. To estimate the total
population of these faint AGN (fAGN hereafter) we adopt a value for
the radio loud fraction derived by \cite{Jiang2007} of 10\%, which has
a scatter of only $\pm$\,4\% with increasing redshift. Hence the
surface density of fAGN is on the order of 310\,deg$^{-2}$, and the
space densities of fAGN and the SMBH mass density arising from this
population are to be multiplied with 10. Since all further
  calculations must be regarded as ``order of magnitude'' estimates,
  we do not specify any errors. IFRS are a new class of object
with particularly extreme radio properties, which may further increase
the scatter in the relation found by \cite{Jiang2007}, hence the
error reported by \citeauthor{Jiang2007} is likely to be too small for our
calculations. Nevertheless, we adopt the main result by
  \cite{Jiang2007} that about one tenth of all AGN across the entire
  investigated redshift range (z$<$4) are radio loud, and we infer the
  number density of fAGN traced by IFRS by simply correcting for this
  factor. Accordingly, the number density of fAGN in this redshift
  range and their SMBH mass density are, given the average mass of
  their SMBHs as discussed above:

\begin{eqnarray}
  n_{fAGN}\,&=&\,10\cdot n_{IFRS}\,=\,1.44\cdot 10^{-5}\,\rm{Mpc}^{-3}\\
  \label{fAGN}
  \rho_{fAGN}\,&=&n_{fAGN}\cdot M_{IFRS}\,\leq\,1.4\cdot 10^3\,\rm{M_{\sun}\,Mpc}^{-3}
\end{eqnarray}

Based on X-ray 0.5-2.0\,keV band observations obtained with both
XMM-{\it Newton} and {\it Chandra}, \cite{Hasinger2005} give a
redshift-dependent space density of AGN for several X-ray
luminosities. For their two faintest categories they find values
around $2\cdot 10^{-5}\,$Mpc$^{-3}$ which, given the many
uncertainties in our estimate, is in very good agreement with the fAGN
number density. Since X-ray surveys are thought to yield a
  reliable census of AGN \citep[see e.g. ][ for an investigation of
the completeness in the {\it Chandra} Deep Fields]{Bauer2004}, this
value should not exceed $5\cdot 10^{-5}\,$Mpc$^{-3}$ given the current
X-ray survey sensitivity limits of
$\sim$\,10$^{-19}$\,W\,m$^{-2}$. Nevertheless there are a few bright
radio sources that do not have an X-ray counterpart even in the 2\,Ms
{\it Chandra} exposure \citep[e.g.][]{Snellen2001} and actual number
counts vary from survey to survey. For example, the XMM-{\it Newton}
survey of the COSMOS field \citep{Brusa2009} yields a space density of
only $10^{-6}\,\rm{Mpc}^{-3}$ at z\,=\,4 whilst the \cite{Gilli2007}
model predicts a value that is four times higher. Therefore we propose to use the IFRS population to complement the number of X-ray-unidentified AGN in the Universe \citep[For a concise review of AGN identification techniques see][]{Mushotzky2004}.

We stress that comparing our results to other surveys must be
  done with exceptional care because the IFRS sample presented here
  may be biased, e.g. towards dusty sources. Further observations of a
  subsample of IFRS with the {\it Herschel} Space Observatory will
  clarify the dust content of these extreme objects.

For AGN at even higher redshifts (z$>$4), \cite{Brandt2005} give a
lower limit on the surface density of AGN of
30-150\,deg$^{-2}$. This value is on the same order of magnitude as
our estimate of the fAGN surface density of $\sim$\,310\,deg$^{-2}$
(ten times the IFRS surface density). Considering that approximately
half of these objects are likely to be situated at such high
redshifts\footnote{With ultra-deep {\it Spitzer} data,
  \cite{Huynh2010} only found counterparts for two out of four IFRS,
  hence they conclude that this IR-undetected AGN exhibiting really
  extreme ratios of radio to infrared flux are located at similar
  extreme redshifts in the range of z\,=\,5-6.}, our estimates are in
good agreement with the \citeauthor{Brandt2005} value.

\subsection{... on the Cosmic X-ray Background}

Since no IFRS in the ELAIS-S1 field is detected as discrete X-ray
source by the corresponding XMM-{\it Newton} survey carried out by
\cite{Puccetti2006}, the question arises where their X-ray radiation
goes. We propose that the X-ray emission contributes to the cosmic
X-ray background (CXB).  Given the surface density of fAGN of
$\sim$\,310\,deg$^{-2}$, and an average mass of their central
  SMBHs of $10^8\,$M$_{\sun}$ one can calculate their
contribution. Following \cite{Dijkstra2004}, an SMBH at its Eddington
limit \citep[we assume accretion at the Eddington limit because the
general trend for high-mass SMBHs indicates this,
see][]{Willott2010} emits a flux density in the observer
0.5-2.0\,keV band of:

\begin{equation}
  S_{SMBH}\,=\,3.8\cdot 10^{-23}\left(\frac{M_{SMBH}}{10^4\,M_{\sun}}\right)\left(\frac{f_{X}}{0.03}\right)\left(\frac{10}{z_{AGN}}\right)^{9/4}\,\rm{W\,m}^{-2}\label{eq:SSMBH}
\end{equation}

Here $f_X$ denotes the fraction of the total energy radiated in the
observer 0.5-2.0\,keV band, its value of 0.03 is taken from the
template spectrum of \cite{Sazonov2004} and will be adopted for all
further calculations. For the fAGN population this implies a
contribution of $CXB_{fAGN}\,=\,9\cdot
10^{-16}$\,W\,m$^{-2}$\,deg$^{-2}$ while assuming an average redshift
of $z_{fAGN}\,=\,4$.

This is in reasonable agreement with the CXB measurements by
\cite{Moretti2003}, who give a total soft (0.5-2.0\,keV) CXB of
$75.3\cdot 10^{-16}$\,W\,m$^{-2}$\,deg$^{-2}$ with a $3.5\cdot
10^{-16}$\,W\,m$^{-2}$\,deg$^{-2}$ component (4.6\,\%) not accounted
for by either point sources, diffuse emission, or scattering. We note
that $CXB_{fAGN}$ is an upper limit for the contribution of fAGN to
the CXB. First, because the adopted mass of their central SMBH is an
upper limit, and second, no dust extinction of the X-ray emission
(predominantly affecting the soft band) was considered, which
would result in the absorption of X-ray flux and re-emission at longer
wavelengths. To calculate the contribution of fAGN to the hard
  CXB, one first needs to integrate the \cite{Sazonov2004} template
  spectrum in the observer's frame 2.0-10\,keV range to obtain a
  fraction for the amount of energy radiated at these wavelengths. We
  derived a fraction of $f_X\,=\,0.1$ which we then used in
  Eqn.~\ref{eq:SSMBH} to calculate the hard band X-ray flux in the
  same manner as done above for the soft band. This results in a
contribution of fAGN to the 2.0-10\,keV region of the CXB of $3\cdot
10^{-15}$\,W\,m$^{-2}$\,deg$^{-2}$, which is about 18\% of the total
value in this band given by \cite{Moretti2003}. These authors also
give 88.8\% as the fraction of resolved point-like and extended
sources contributing to the hard fraction of the CXB, implying an
unresolved component of 11.2\%. So also in the hard band does the fAGN
population nicely account for the unresolved fraction of the
CXB. However, even though our estimates match other observations and
predictions, these calculations should be understood only as a rough
estimate of the IFRS number densities and hence fAGN population
contributing to the CXB.

\subsection{... on Structure Formation}

The presence of a population of AGN-driven objects of which at least a
fraction (up to 50\%) is likely to be located at very high redshifts
(z\,$>$\,5) puts several constraints on the formation scenario of
SMBHs shortly after the Big Bang. The Millennium Simulation
\citep{Springel2005}, to date the largest cosmological simulation
probing $\Lambda$CDM cosmology, contains only one massive
($M_{DM-halo}\,=\,5.5\cdot 10^{12}\,$M$_{\sun}$) halo at z\,=\,6.2
which is a candidate for a quasar sufficiently bright to be
observed by the SDSS. The SMBH mass density at z\,=\,6 of
$\sim\,10^{-9}$\,M$_{\sun}$\,Mpc$^{-3}$ derived from this simulation
is therefore several orders of magnitude lower than the mass density
we extrapolate from our fAGN (Eq.~\ref{fAGN}). We note, however, that
bright quasars are still very rare objects which can be treated as
statistical ``spikes'' in the primordial density profile and thus do
not very much affect current theories concerning structure formation.
Given the very small number density of such bright SDSS quasars
  \citep[about 0.012\,deg$^{-2}$, see][]{Schneider2010} or the
  powerful high redshift radio sources \citep[which need to be
    selected from all-sky radio surveys, see][]{Seymour2007}, we point
  out that IFRS resp. fAGN can play an important role just
  because of their abundance of several tens to several hundred per
  square degree.

One can also compare the mass densities to models of SMBH
  formation. \cite{Bromley2004} assert that major mergers account for
most of the high-z SMBHs. Their approach yields a close match to
our mass density, in particular when their simulation scenario G is
considered, where the probability of forming a seed black hole during
a merger event is less than unity and the fraction of halo gas
accreted onto the black hole is constant. This scenario yields a mass
density of $\sim 10^3$\,M$_{\sun}$\,Mpc$^{-3}$ when SMBHs with
$10^8$\,M$_{\sun}$ at z\,=\,6 are considered. Since not all fAGN are
likely to be located at such high redshifts, their model variant B, in
which the halo gas accretion fraction is proportional to the halo
virial velocity squared, is also promising, and yields a mass density
of $\sim 5\cdot 10^2$\,M$_{\sun}$\,Mpc$^{-3}$. Another attempt to
model SMBH properties at such high redshifts was carried out by
\cite{Tanaka2009} using both accretion and merger scenarios. They find
that to produce sufficient high-mass SMBHs that power the most
luminous SDSS quasars ($\geq\,10^9$\,M$_{\sun}$), their models yield a
mass density of $10^5$\,M$_{\sun}$\,Mpc$^{-3}$ for SMBHs with masses
around $10^8$\,M$_{\sun}$, emphasising that this is too large by a
factor of 100 to 1000. Hence their estimate for such
$10^8$\,M$_{\sun}$ SMBHs is of the order
$10^2$\,M$_{\sun}$\,Mpc$^{-3}$, too.

We conclude that a non-negligible fraction of lower-mass SMBHs at the
highest redshifts can put constraints on structure formation
after the Big Bang -- in particular considering SMBH growth
models connected to the build-up of galaxies. Most of these
theoretical frameworks focus only on the high-mass SMBHs necessary to
power the most luminous quasars. Present observations of lower-mass
SMBHs are not possible using currently available instruments with
reasonable integration times, but radio observations can play a
crucial role to investigate the effects of black holes on various
scales in the early universe.

\section{Summary, Conclusions and Outlook}

We have investigated the implications of Infrared-Faint Radio Sources
(IFRS) being high-redshift AGN. We draw consequences on AGN number
counts and SMBH mass densities, the Cosmic X-ray Background (CXB) and
theoretical models of structure formation after the Big Bang. Our main
results are:

\begin{enumerate}

\item The IFRS surface density, which is strongly tied to the
  sensitivity limit of the parent radio survey (fig. \ref{IFRS-dens}),
  can be extrapolated to a ``zero noise'' survey, yielding an
  intrinsic surface density of (30.8$\pm$ 15.0)\,deg$^{-2}$. Previous
  work indicates that IFRS are AGN-driven objects at high redshifts,
  hence this surface density can be converted to a space density of
  $n_{IFRS}\,=\,1.44\cdot 10^{-6}$\,Mpc$^{-3}$ if they are in the
  redshift range 3\,$\leq$\,z\,$\leq$\,6.

\item Correcting for the finding that $\sim$\,10\% of all AGN are
  radio-loud, a population of high-redshift AGN (``fAGN'') is traced
  by IFRS, of which 10\,\% are visible only in the radio. Almost no
  IR, optical, and X-ray detections have been made so far. This can be
  used to constrain the lower-luminosity population of high-redshift
  AGN. Comparing the number density of these fAGN,
  $n_{fAGN}\,=\,1.44\cdot 10^{-5}\,\rm{Mpc}^{-3}$, to number densities
  of low-luminous AGN obtained in the X-ray regime we found them to be
  in good agreement ($n_{X-ray}\,=\,2\cdot 10^{-5}\,$Mpc$^{-3}$). For
  the most distant AGN (z\,$>$\,4), the X-ray-derived lower limit of
  30-150\,deg$^{-2}$ is more than doubled by the fAGN, which have a
  surface density on the order of 310\,deg$^{-2}$.

\item We find that fAGN can account for the missing unresolved
  components in both the soft (0.5-2.0\,keV) and hard (2.0-10\,keV)
  band of the CXB. Our estimates yield $CBX_{fAGN,
    soft}\,=\,9\cdot 10^{-16}$\,W\,m$^{-2}$\,deg$^{-2}$ and
  $CBX_{fAGN, hard}\,=\,3\cdot
  10^{-15}$\,W\,m$^{-2}$\,deg$^{-2}$. These values represent 9\% of
  the soft CXB component (cf. to $\sim$\,4.6\,\% which are regarded as
  the missing unresolved fraction), and 18\% of the hard CXB component
  (cf. to $\sim$\,11.2\% which are regarded as the missing unresolved
  fraction).

\item The existence of a non-negligible fraction of lower-mass SMBHs
  at such high redshifts is in good agreement with recent cosmological
  simulations, favouring a formation scenario for early SMBHs based on
  halo gas accretion as well as major mergers. Only considering
  statistical ``spikes'' in the primordial density fluctuations as the
  ancestors of the SMBH population in the early universe is not
  sufficient. Our findings contrast older simulations
    which explain the low number of SMBHs with the low number of
  high-redshift quasars known, an argument which has now become
    invalid.
\end{enumerate}

Future observations of the IFRS population will further constrain the
nature and thus the astrophysical significance of these objects. In
particular sub-mm observatories such as ESO's ALMA array or the {\it
  Herschel} Space Observatory will be able to provide constraints on
the MIR/FIR emission of IFRS. They will clarify if the IFRS
population predominantly consists of highly obscured type II
AGN, or if IFRS in general show flatter SEDs without any distinct
far-infrared bump produced by dust. Corresponding deep {\it Herschel}
observations have already been awarded time and will be executed in
the forthcoming observational period. They will put constraints
on the star formation history of the universe by estimating dust
masses for objects at redshifts potentially as high as z\,=\,6.

\begin{acknowledgements}
  This research has made use of the NASA/IPAC Extragalactic Database
  (NED) which is operated by the Jet Propulsion Laboratory, California
  Institute of Technology, under contract with the National
  Aeronautics and Space Administration.
\end{acknowledgements}

\bibliographystyle{aa}
\bibliography{IFRS-dens_bib}

\begin{table*}[t]
    \caption{Catalog of known IFRS in both ATLAS fields (CDF-S and ELAIS-S1), the {\it Spitzer} extragalactic First Look Survey (xFLS), and the COSMOS field.}
    \label{IFRS-cat}
    \centering
    \begin{tabular}{llccccccc}
    \hline
    \hline
    \noalign{\smallskip}
    Survey & Identifier\tablefootmark{a} & RA (J2000) & Dec (J2000) & $S_{1.4\,GHz}$ & $\Delta S$ & $S_{20\,\rm{cm}}/S_{3.6\,\rm{\mu m}}$ & SNR & flag\tablefootmark{b} \\
    &  & h:m:s & d:m:s & mJy & mJy & & & \\
    \noalign{\smallskip}
    \hline
    \noalign{\smallskip}
    ATLAS/ELAIS-S1 & S1021 & 00:32:55.534 & -43:16:27.15 & 16.1 & 0.81 & 575 & 19 & \\
    ATLAS/ELAIS-S1 & S645 & 00:39:34.763 & -43:42:22.58 & 5.46 & 0.4 & 780 & 13 & \\
    ATLAS/ELAIS-S1 & S1018 & 00:29:46.525 & -43:15:54.52 & 27.34 & 1.37 & 1012 & 19 & \\
    ATLAS/ELAIS-S1 & S5 & 00:37:09.365 & -44:43:48.11 & 15.16 & 0.76 & 1082 & 19 & \\
    ATLAS/ELAIS-S1 & S1239 & 00:35:47.969 & -42:56:55.40 & 23.19 & 1.16 & 1220 & 19 & \\
    ATLAS/ELAIS-S1 & S66 & 00:39:42.452 & -44:27:13.77 & 33.58 & 1.71 & 1865 & 19 & \\
    ATLAS/ELAIS-S1 & S1156 & 00:36:45.856 & -43:05:47.39 & 31.77 & 1.59 & 2888 & 19 & \\
    ATLAS/ELAIS-S1 & S11 & 00:32:07.444 & -44:39:57.18 & 1.67 & 0.014 & 557 & 14 & \\
    ATLAS/ELAIS-S1 & S419 & 00:33:22.766 & -43:59:15.37 & 1.67 & 0.010 & 557 & 15 & \\
    ATLAS/ELAIS-S1 & S798 & 00:39:07.934 & -43:32:05.83 & 7.79 & 0.43 & $>$2597 & 18 & M\\
    ATLAS/ELAIS-S1 & S749 & 00:29:05.229 & -43:34:03.94 & 7.01 & 0.41 & $>$2337 & 17 & M\\
    ATLAS/ELAIS-S1 & S973 & 00:38:44.139 & -43:19:20.43 & 9.14 & 0.46 & $>$3046 & 19 & M\\
    ATLAS/ELAIS-S1 & S427 & 00:34:11.592 & -43:58:17.04 & 21.36 & 1.07 & $>$7120 & 19 & M\\
    ATLAS/ELAIS-S1 & S509 & 00:31:38.633 & -43:52:20.80 & 22.2 & 1.11 & $>$7400 & 20 & M\\
    ATLAS/ELAIS-S1 & S201 & 00:31:30.068 & -44:15:10.69 & 5.05 & 0.25 & $>$1683 & 20 & M\\
    ATLAS/CDF-S & S114 & 03:27:59.894 & -27:55:54.73 & 7.2 & 0.042 & $>$2400 & 38 & N\\
    ATLAS/CDF-S & S194 & 03:29:28.594 & -28:36:18.81 & 6.1 & 0.064 & $>$2033 & 32 & N\\
    ATLAS/CDF-S & S703 & 03:35:31.025 & -27:27:02.20 & 26.1 & 0.040 & $>$8700 & 140 & N\\
    ATLAS/CDF-S & S292 & 03:30:56.949 & -28:56:37.29 & 22.3 & 0.081 & 1842 & 119 & \\
    ATLAS/CDF-S & S649 & 03:34:52.846 & -27:58:13.05 & 5.7 & 0.034 & 1838 & 30 & \\
    ATLAS/CDF-S & S618 & 03:34:29.754 & -27:17:44.95 & 42.5 & 0.059 & 1660 & 228 & \\
    ATLAS/CDF-S & S574 & 03:33:53.279 & -28:05:07.31 & 11.9 & 0.029 & 1091 & 63 & \\
    ATLAS/CDF-S & S94 & 03:27:40.727 & -28:54:13.48 & 8.9 & 0.08 & 801 & 47 & \\
    ATLAS/CDF-S & S603 & 03:34:13.759 & -28:35:47.47 & 13.2 & 0.039 & 709 & 70 & \\
    ATLAS/CDF-S & S713 & 03:35:37.525 & -27:50:57.88 & 16.4 & 0.069 & 643 & 88 & \\
    ATLAS/CDF-S & S539 & 03:33:30.542 & -28:54:28.22 & 9.1 & 0.070 & 640 & 48 & \\
    ATLAS/CDF-S & S265 & 03:30:34.661 & -28:27:06.51 & 18.6 & 0.034 & 634 & 100 & \\
    ATLAS/CDF-S & S97 & 03:27:41.700 & -27:42:36.61 & 4.3 & 0.033 & 614 & 23 & \\
    ATLAS/CDF-S & S520 & 03:33:16.754 & -28:00:16.02 & 3.7 & 0.020 & 500 & 19 & \\
    xFLS & 2009 & 17:18:56.948 & +60:21:11.84 & 14.111 & 0.599 & $>$1568 & 23 & \\
    xFLS & 2779 & 17:22:30.252 & +60:08:49.91 & 6.582 & 0.33 & $>$731 & 19 & \\
    xFLS & 2371 & 17:20:37.485 & +60:14:42.80 & 5.34 & 0.228 & $\sim$500 & 23 & s\\
    xFLS & 1050 & 17:14:36.539 & +59:44:56.44 & 4.058 & 0.174 & $\sim$500 & 23 & s\\
    xFLS & 1120 & 17:14:53.815 & +59:43:29.28 & 3.685 & 0.158 & $\sim$500 & 23 & s\\
    xFLS & 478 & 17:11:48.526 & +59:10:38.87 & 35.815 & 1.52 & 1831 & 23 & uc\\
    xFLS & 600 & 17:12:25.462 & +59:40:14.98 & 16.2 & 0.688 & 730 & 23 & uc\\
    xFLS & 760 & 17:13:15.445 & +59:03:02.66 & 2.47 & 0.11 & $>$798 & 25 & GA\\
    xFLS & 929 & 17:14:03.986 & +59:53:17.54 & 2.42 & 0.10 & $>$780 & 24 & GA\\
    xFLS & 1050 & 17:14:36.539 & +59:44:56.44 & 4.06 & 0.17 & $>$1310 & 24 & GA\\
    xFLS & 1120 & 17:14:53.815 & +59:43:29.28 & 3.69 & 0.16 & $>$1190 & 23 & GA\\
    xFLS & 2398 & 17:20:46.548 & +58:47:28.91 & 3.58 & 0.15 & $>$1155 & 22 & GA\\
    xFLS & 2928 & 17:23:15.537 & +59:09:19.46 & 1.41 & 0.06 & $\sim$500 & 21 & GA,s\\
    COSMOS & J095957.97+020109.7 & 09:59:57.97 & +02:01:09.7 & 7.070 & 0.099 & 542 & 643 & \\
    COSMOS & J095959.16+014837.8 & 09:59:59.16 & +01:48:37.8 & 9.600 & 0.110 & 886 & 873 & \\
    COSMOS & J100058.05+015129.0 & 10:00:58.05 & +01:51:29.0 & 11.950 & 0.120 & 862 & 1195 & \\
    COSMOS & J100109.28+021721.7 & 10:01:09.28 & +02:17:21.7 & 4.100 & 0.010 & 522 & 410 & \\
    COSMOS & J100120.06+023443.7 & 10:01:20.06 & +02:34:43.7 & 10.796 & 0.112 & 1301 & 830 & \\
    COSMOS & J100209.06+021602.6 & 10:02:09.06 & +02:16:02.6 & 6.136 & 0.068 & 515 & 558 & \\
    COSMOS & J100212.06+023134.8 & 10:02:12:06 & +02:31:34.8 & 18.690 & 0.200 & 649 & 1699 & \\
    COSMOS & J100114.12+015444.3 & 10:01:14.12 & +01:54:44.3 & 7.000 & 0.044 & 538 & 636 & \\
    COSMOS & J095803.21+021357.7 & 09:58:03.21 & +02:13:57.7 & 25.302 & 0.264 & 1438 & 904 & \\
    COSMOS & J100129.35+014027.1 & 10:01:31.10 & +01:40:18.0 & 6.330 & 0.597 & 1476 & 452 & \\
    COSMOS & J095908.87+013603.6 & 09:59:08.87 & +01:36:03.6 & 6.484 & 0.085 & 500 & 359 & \\
    COSMOS & J100252.88+015549.7 & 10:02:25.88 & +01:55:49.7 & 17.543 & 0.184 & 931 & 975 & \\
    COSMOS & J095838.01+013217.1 & 09:58:38.01 & +01:32:17.1 & 4.577 & 0.095 & 509 & 254 & \\
    \noalign{\smallskip}
  \hline
  \end{tabular}
  \tablefoot{
  \tablefoottext{a}{The identifiers are given as in the original publications to ensure easy data handling.}\\
  \tablefoottext{b}{
s: uncertain IFRS (e.g. because the IR data is too shallow to yield a clear counterpart detection with reliable flux measurements, so that the error in IR flux is large).
uc: uncataloged counterpart.
GA: IFRS in \cite{Garn2008}.
M: IFRS in \cite{Middelberg2008a}.
N: IFRS in \cite{Norris2006}.}}
\end{table*}

\end{document}